\documentclass[12pt]{article}
\usepackage{pgfplots}
\usetikzlibrary{decorations.pathmorphing}

\tikzset{snake it/.style={decorate, decoration=snake}}
\pgfplotsset{compat=1.10}
\usepgfplotslibrary{fillbetween}
\usetikzlibrary{patterns}
\usepackage{draft} 

\usepackage{hyperref}
\usepackage{graphicx,color,subfigure}
\usepackage{cite}
\usepackage{mciteplus}
\usepackage{skak}
\usepackage{xcolor}
\usepackage{empheq}
\usepackage{tikz}
\DeclareFontFamily{OT1}{pzc}{}
\DeclareFontShape{OT1}{pzc}{m}{it}{<-> s * [1.10] pzcmi7t}{}
\DeclareMathAlphabet{\mathpzc}{OT1}{pzc}{m}{it}

\def\be#1\ee{\begin{align}#1\end{align}}



\begin{document}

\unitlength = .8mm

\begin{titlepage}

\begin{center}

\hfill \\
\hfill \\
\vskip 1cm

\title{\Huge Horowitz-Polchinski Solutions at Large $k$}

\author{Jinwei Chu and David Kutasov}
\address{
Leinweber Institute for Theoretical Physics,\\  Enrico Fermi Institute, and Department of Physics\\ University of Chicago, Chicago IL 60637
}
\vskip 1cm

\email{jinweichu@uchicago.edu, dkutasov@uchicago.edu}

\end{center}

\abstract{In \cite{Chu:2025boe}, we introduced an approximation that allows one to study Horowitz-Polchinski backgrounds beyond the weak coupling regime. In this paper we describe the resulting solutions, and discuss a few related issues. 
}

\vfill

\end{titlepage}

\eject

\begingroup
\hypersetup{linkcolor=black}
\tableofcontents
\endgroup

\vfill
\eject

\section{Introduction}

String theory on ${\mathbb{R}}^d\times {\mathbb{S}}^1$ plays an important role in studying  thermodynamics in $d+1$ spacetime dimensions. The ${\mathbb{S}}^1$ corresponds to Euclidean time; its circumference $\beta$ is the inverse temperature. Of particular interest is the region $\beta\sim\beta_H$, which corresponds to temperatures near the Hagedorn temperature \cite{Polchinski:1998rq}. This region is relevant for the string-black hole correspondence \cite{Horowitz:1996nw}, and for the description of small black holes in string theory. 

Horowitz-Polchinski (HP) solutions \cite{Horowitz:1997jc} (see \cite{Chen:2021dsw,Balthazar:2022szl,Balthazar:2022hno,Mazel:2024alu,Bedroya:2024igb} for more recent work on these and related solutions)  are Euclidean backgrounds that depend on the radial coordinate in ${\mathbb{R}}^d$, $r$, and asymptote at large $r$ to ${\mathbb{R}}^d\times {\mathbb{S}}^1$, with $\beta\sim\beta_H$. They are qualitatively similar to small Euclidean Schwarzschild black holes, in the sense that both backgrounds preserve the $SO(d)$ rotation symmetry, and break the winding symmetry around the Euclidean time circle. However, there are some differences as well, and the precise relation between the two classes of backgrounds remains an open problem. 

In the original work \cite{Horowitz:1997jc}, and the more recent \cite{Balthazar:2022hno}, it was shown that for $d\le 6$, HP solutions can be described in terms of an effective field theory (EFT) on ${\mathbb{R}}^d$ that contains two fields, the radion $\phi(x)$ that parametrizes the radius of the Euclidean time circle, and the winding tachyon $\chi(x)$. This EFT is reliable near the Hagedorn temperature, and gives a calculable contribution to the free energy.

For $d>6$, the EFT of \cite{Horowitz:1997jc,Balthazar:2022hno} breaks down. As shown in \cite{Balthazar:2022hno}, in order to study the HP backgrounds in this regime, one must include in the effective Lagrangian terms of arbitrary order in fields and derivatives. This is a reflection of the fact that the corresponding worldsheet CFT is in general strongly coupled. 

In a recent paper \cite{Chu:2025boe}, we proposed an approach to circumventing this problem. Our proposal is based on the observation that the theory that gives rise to the HP solutions has an underlying $SU(2)_L\times SU(2)_R$ symmetry. This symmetry is most apparent at the Hagedorn temperature, $\beta=\beta_H$, but we showed that it is very useful for organizing the dynamics away from the Hagedorn temperature as well. On the worldsheet, this symmetry is described by an affine Lie algebra with a level of order one (one for the bosonic string and two for the superstring \cite{Balthazar:2022szl}). In \cite{Chu:2025boe} we pointed out that if we continue this level to a large value, $k\gg 1$, the problem simplifies. In particular, it can again be described by an EFT.  One can hope that studying this large $k$ EFT may provide insight into the problem of interest, which has small $k$, in the spirit of large $N$ approximations in QFT. 

One of the main results of \cite{Chu:2025boe} was the derivation, following \cite{Kutasov:1989dt} (see also \cite{Cvetic:2000dm,Sagkrioti:2018abh}), of the effective Lagrangian that governs the large $k$ theory. It is given (up to an overall multiplicative constant), by 
\ie
\label{resclagdil}
L_{\rm eff}=\sqrt{g}e^{-2\Phi}\left[-\mathcal{R}-4(\nabla\Phi)^2+L_K+ V(\phi,\chi,\chi^*)\right]\ .
\fe
Here $\phi$ and $\chi$ are as above, $\Phi$ is the $d$ dimensional dilaton, and $g_{ij}$ the metric on ${\mathbb{R}}^d$. The kinetic and potential terms are given by
\begin{equation}
\label{kinterm}
    L_K=\frac{(2\pi)^2|\nabla \chi|^2}{(1-2\pi^2|\chi|^2)^2}+\frac{(2\pi)^2(\nabla \phi)^2}{(1-4\pi^2\phi^2)^2}\ ,
\end{equation}
and 
\begin{equation}
\label{Vchiphi}
    V(\phi,\chi,\chi^*)=\frac{4}{\alpha'}\left(\frac{4\pi^2|\chi|^2}{(1-2\pi\phi)(1-2\pi^2|\chi|^2)^2}-\frac{1}{(1-2\pi^2|\chi|^2)^2}+1\right)~,
\end{equation}
respectively. As discussed in~\cite{Chu:2025boe}, the coordinates on $\mathbb{R}^d$ have been rescaled to
\begin{equation}
\label{resx}
    x^i\to \sqrt{\frac{k}{2}}x^i\ ,\quad i=1,2,\cdots,d\ ,
\end{equation}
so that the e.o.m. are independent of $k$.

The main goal of this note is to study the solutions of the Euler-Lagrange equations of \eqref{resclagdil} -- \eqref{Vchiphi}, with the HP boundary conditions. We will also comment on the analogs of black hole solutions in this setup. This note is a companion paper to \cite{Chu:2025boe}, and we refer the reader to that paper for further background and discussion of the HP problem, as well as additional references.

As explained above, we are interested in spherically symmetric solutions. For such solutions, we can choose a parametrization in which the string frame metric on ${\mathbb{R}}^d$ takes the form 
\begin{equation}
\label{dsr}
    ds^2=e^{2h(r)}dr^2+r^2d\Omega_{d-1}^2\ .
\end{equation}
The scalar curvature of this metric is
\begin{equation}
    \mathcal{R}(r)=\frac{(d-1)(d-2)}{r^2}(1-e^{-2h(r)})+\frac{d-1}{r}\frac{2h'(r)}{e^{2h(r)}}\ .
\end{equation}
Substituting it into the Lagrangian (\ref{resclagdil}) yields a one-dimensional theory of four fields, $h$, $\Phi$, $\chi$ and $\phi$. The Lagrangian (\ref{resclagdil}) takes the form
\ie
\label{Lr}
L_{\rm eff}=r^{d-1}e^{h(r)-2\Phi(r)}\left[-\mathcal{R}(r)-4e^{-2h(r)}(\Phi'(r))^2+e^{-2h(r)}\tilde L_K+ V(\phi,\chi,\chi^*)\right]\ ,
\fe
where
\begin{equation}
\label{LK}
   \tilde  L_K=\frac{(2\pi)^2|\chi'(r)|^2}{(1-2\pi^2|\chi(r)|^2)^2}+\frac{(2\pi)^2( \phi'(r))^2}{(1-4\pi^2\phi^2(r))^2}\ .
\end{equation}

The equation of motion for $h(r)$ is algebraic,
\ie
(d-1)(d-2)(1-e^{2h})+r^2\left(4\Phi'^2-4\frac{d-1}{r}\Phi'-\tilde L_K\right)+r^2e^{2h}V(\phi,\chi,\chi^*)=0\ .
\fe
It can be used to express $h(r)$ in terms of the other fields, 
\ie
\label{eh}
    e^h=\sqrt{\frac{(d-1)(d-2)+r^2\left(4\Phi'^2-4\frac{d-1}{r}\Phi'-\tilde L_K\right)}{(d-1)(d-2)-r^2V(\phi,\chi,\chi^*)}}\ .
\fe
The e.o.m. for $\Phi$, $\chi$ and $\phi$  are second-order differential equations:
\ie
\label{aa}
\frac{d}{dr}(4r^{d-1}e^{-h-2\Phi}\Phi')=r^{d-1}e^{h-2\Phi}\left[-\mathcal{R}-4e^{-2h}(\Phi')^2+e^{-2h}\tilde L_K+ V(\phi,\chi,\chi^*)\right]~,
\fe
\ie
\label{bb}
\frac{d}{dr}\left(r^{d-1}e^{-h-2\Phi}\frac{(2\pi)^2\chi'}{(1-2\pi^2|\chi|^2)^2}\right)=&r^{d-1}e^{h-2\Phi}\left[e^{-2h}\frac{(2\pi)^4|\chi'|^2\chi}{(1-2\pi^2|\chi|^2)^3}+\frac{\partial V}{\partial \chi^*}\right]~,
\fe
\ie
\label{cc}
\frac{d}{dr}\left(r^{d-1}e^{-h-2\Phi}\frac{2(2\pi)^2\phi'}{(1-4\pi^2\phi^2)^2}\right)=&r^{d-1}e^{h-2\Phi}\left[e^{-2h}\frac{2^6\pi^4(\phi')^2\phi}{(1-4\pi^2\phi^2)^3}+\frac{\partial V}{\partial \phi}\right]~,
\fe
where, from (\ref{Vchiphi}),
\ie
\label{dd}
\frac{\partial V}{\partial \chi^*}=\frac{32\pi^3(\phi+\pi|\chi|^2)\chi}{\alpha'(1-2\pi\phi)(1-2\pi^2|\chi|^2)^3}\ ,\quad \frac{\partial V}{\partial \phi}=\frac{32\pi^3|\chi|^2}{\alpha'(1-2\pi\phi)^2(1-2\pi^2|\chi|^2)^2}\ .
\fe
In the next section we solve these equations numerically. As explained in \cite{Chu:2025boe}, we can take $\chi$ to be real and positive without loss of generality, and we will do so below. 

\section{Numerics}

Solving equations (\ref{eh}) -- (\ref{cc}) numerically requires six boundary conditions. For studying HP-like solutions, it is convenient to impose them at $r=0$:
\begin{equation}
\label{bc}
    \Phi(0)=0\ ,\quad \chi(0)=\chi_0\ ,\quad \phi(0)=\phi_0\ ,\quad \Phi'(0)=\chi'(0)=\phi'(0)=0\ .
\end{equation}
The value of $\Phi(0)$ is arbitrary due to the shift symmetry of the Lagrangian \eqref{Lr}, $\Phi\to\Phi+{\rm const}$. We set it to zero in \eqref{bc}, but the only meaningful quantity is the difference between the values of $\Phi$ at zero and at infinity. The three conditions on the first derivatives of the fields in \eqref{bc} follow from regularity at $r=0$. 

The solutions of this system of equations form a one-parameter family. In the original HP problem, this family is labeled by the temperature, but in our large $k$ approximation it is more convenient to use $\chi_0$ as the parameter. To find the solutions, we fix this parameter to a particular value, and vary $\phi_0$ demanding that $\chi(r)$ is normalizable.  This gives a discrete set of solutions with different $\phi_0$. The lowest of these solutions corresponds to a monotonic $\chi(r)$, and a $\phi(r)$ that approaches a constant $\phi_\infty$ as $r\to\infty$. This constant depends on $\chi_0$, and can be thought of as the large $k$ analog of the parameter that labels the radius of the Euclidean time circle at infinity \cite{Chu:2025boe}.

Before turning to the numerical results, we note that: (1)~as in \cite{Balthazar:2022szl,Balthazar:2022hno}, we will  allow the dimension $d$ to take non-integer values; (2)~to simplify the calculations, we will omit the overall multiplicative constant in the potential \eqref{Vchiphi}. It can be absorbed in a rescaling of $r$; (3)~our main interest will be in the region $d>6$. For $d=6+\epsilon$ with $\epsilon \ll 1$, it was shown in \cite{Balthazar:2022hno} that the fields $\Phi$ and $h$ can be neglected, and the weak field approximation is accurate. Here we will focus on what happens for finite $\epsilon$, where we will use the results of \cite{Chu:2025boe}. 

\begin{figure}
	\centering
	\subfigure[]{
	\begin{minipage}[t]{0.32\linewidth}
	\centering
	\includegraphics[width=2in]{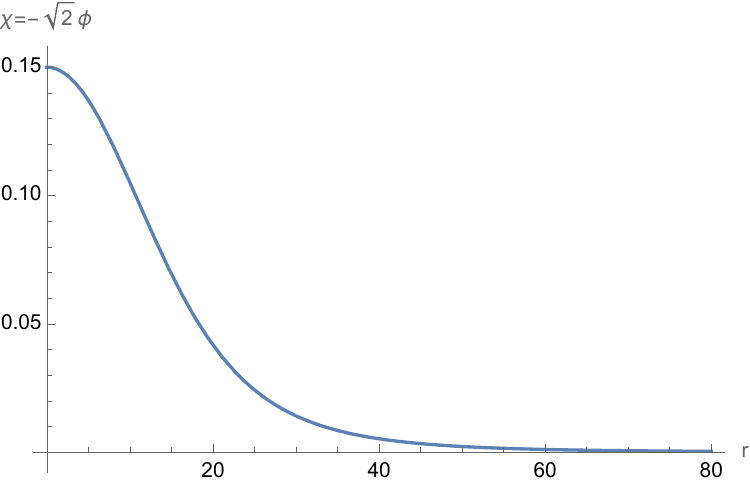}\label{chid65}
	\end{minipage}}
	\subfigure[]{
	\begin{minipage}[t]{0.32\linewidth}
	\centering
	\includegraphics[width=2in]{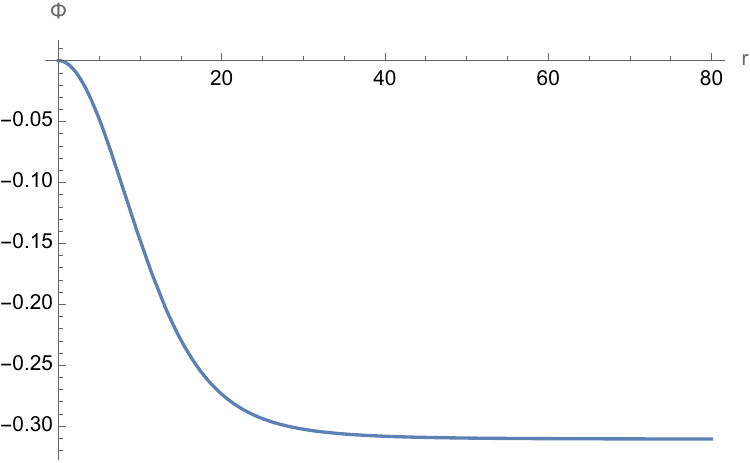}\label{Phid65}
	\end{minipage}}
        \subfigure[]{
	\begin{minipage}[t]{0.32\linewidth}
	\centering
	\includegraphics[width=2in]{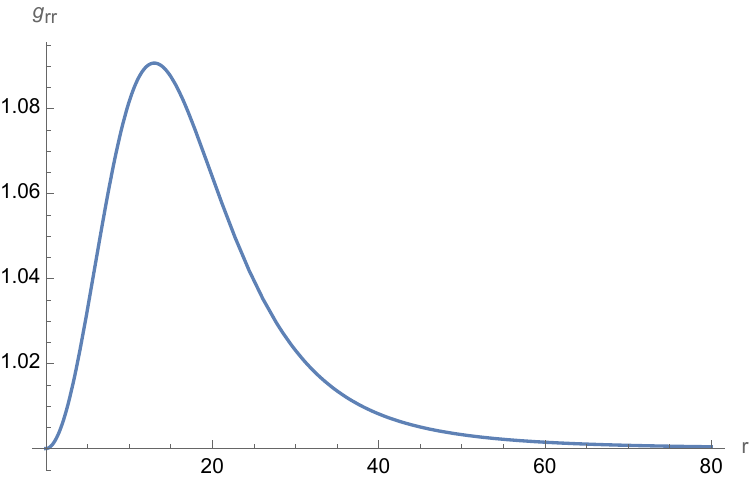}\label{grrd65}
	\end{minipage}}
	\centering
\caption{\label{chiPhid65}The profiles of $\chi=-\sqrt2\phi$, $\Phi$ and $g_{rr}$ for $d=6.5$.}
\end{figure}

In figure \ref{chiPhid65}, we show the numerical solution of (\ref{eh}) -- (\ref{cc}) for $d=6.5$ and $\chi=-\sqrt{2}\phi$. As discussed in \cite{Chu:2025boe}, the latter is the requirement that the $SU(2)_L\times SU(2)_R$ symmetry of the Lagrangian is broken by the solution to a diagonal $SU(2)$. In the HP context, this happens at the Hagedorn temperature, so here we are studying the large $k$ analog of that case. 

The behavior of $\chi(r)$ in figure \ref{chid65} is similar to that in the weak field analyses \cite{Horowitz:1997jc,Chen:2021dsw,Balthazar:2022szl,Balthazar:2022hno,Mazel:2024alu,Bedroya:2024igb}. Figures \ref{Phid65} and \ref{grrd65} show the back-reaction of the dilaton $\Phi$ and the metric $g_{rr}$ on the non-zero $\chi(r)$. As expected, this back-reaction is localized to a region of radial size comparable to the size of the profile $\chi(r)$. It is also relatively muted -- e.g. the value of $g_{rr}$ at the maximum is about 2\% larger than at infinity, in agreement with expectations (that for $d=6+\epsilon$, the back-reaction goes to zero as $\epsilon\to 0$). Another notable feature is that the dilaton at the origin is larger than the one at infinity. This raises the question whether the difference between the two diverges at some value of the dimension. We will address this question below.  

In figure \ref{grrd65} we see that $g_{rr}=1$ at $r=0,\infty$. The latter is due to the fact that $\chi,\phi$ go to zero at infinity, so in this limit the background asymptotes to flat space. The former follows from substituting the boundary conditions (\ref{bc}) into (\ref{eh}), which leads to
\begin{equation}
    g_{rr}(0)=e^{2h(0)}=1\ .
\end{equation}
\begin{figure}
	\centering
	\subfigure[]{
	\begin{minipage}[t]{0.45\linewidth}
	\centering
	\includegraphics[width=2.5in]{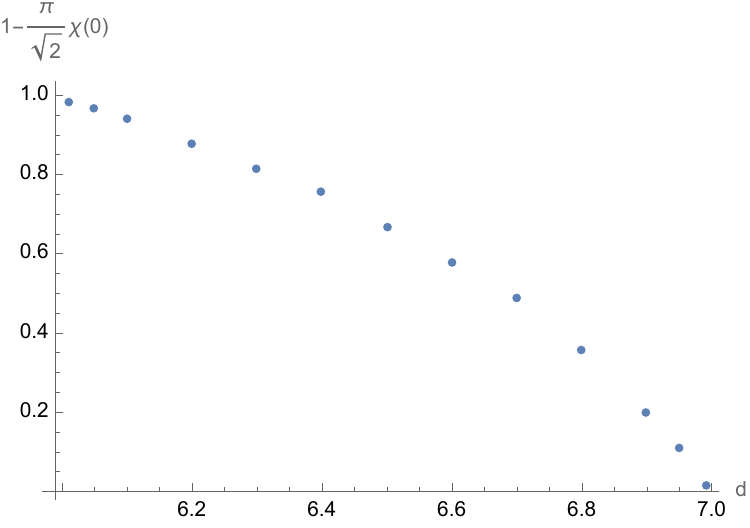}\label{chi0d}
	\end{minipage}}
	\subfigure[]{
	\begin{minipage}[t]{0.45\linewidth}
	\centering
	\includegraphics[width=2.5in]{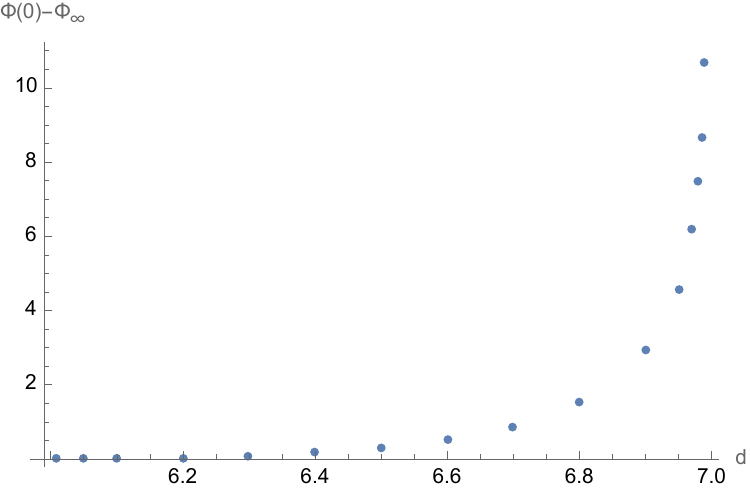}\label{Phid}
	\end{minipage}}
	\centering
\caption{\label{chi0Phid}Dependence on $d$ of the solution with $\chi=-\sqrt{2}\phi$.}
\end{figure}
\ \ An interesting question is what happens to the solution in figure \ref{chiPhid65} when we change the dimension $d$. In figure \ref{chi0Phid} we answer this question for the particular case $\chi=-\sqrt{2}\phi$, in which the solution preserves a diagonal $SU(2)$. Figure \ref{chi0d} shows the dependence of the maximal value of $\chi$, $\chi(0)$, on the dimension. The particular parametrization of the vertical axis in this figure is useful, since the point $\chi=1/\sqrt2\pi$ corresponds to a singularity of the kinetic \eqref{kinterm} and potential \eqref{Vchiphi} terms in the effective Lagrangian.  

We learn from figure \ref{chi0d} that for general $6<d<d_c\approx 7.172$, the solution is regular for all $r$. Geometrically, it describes a deformed three-sphere, whose  deformation increases as $r$ decreases; it is largest at $r=0$. For $d=6+\epsilon$, the deformation is very small for all $r$, in agreement with the results of \cite{Balthazar:2022hno} for the HP problem. We will describe that deformation in the next section. On the other hand, as $d\to d_c$, the deformation grows,  and the solution becomes singular at small $r$.

Figure \ref{Phid} shows the difference between the value of the dilaton at small and large $r$. If we fix the value of the dilaton at infinity, its value at the origin grows without bound as $d\to d_c$. Thus, the analysis breaks down at that value of $d$. For $d>d_c$, the HP-type solutions described above cease to exist.

\begin{figure}
	\centering
	\subfigure[]{
	\begin{minipage}[t]{0.32\linewidth}
	\centering
	\includegraphics[width=2in]{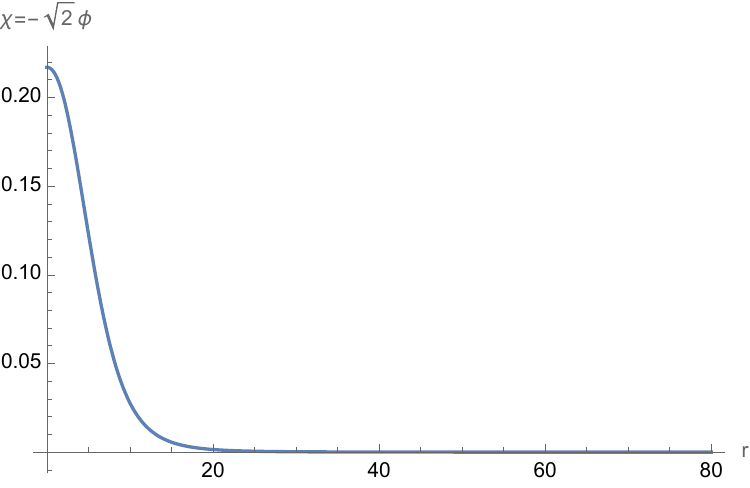}\label{chid7171}
	\end{minipage}}
	\subfigure[]{
	\begin{minipage}[t]{0.32\linewidth}
	\centering
	\includegraphics[width=2in]{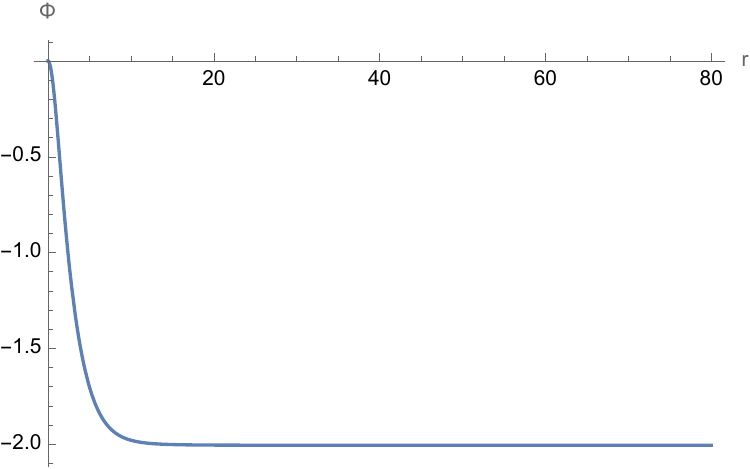}\label{Phid7171}
	\end{minipage}}
    \subfigure[]{
	\begin{minipage}[t]{0.32\linewidth}
	\centering
	\includegraphics[width=2in]{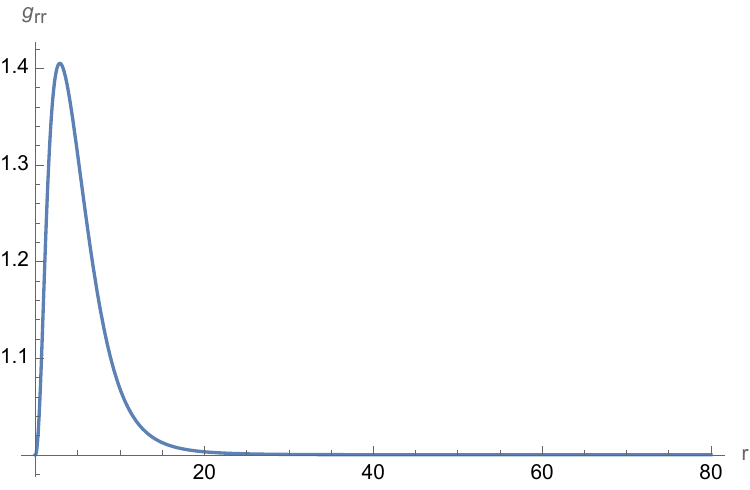}\label{grrd7171}
	\end{minipage}}
	\centering
\caption{\label{d7171}The profiles of $\chi$, $\Phi$ and $g_{rr}$ for $d=7.171$.}
\end{figure}
The curves in figure \ref{chi0Phid} were obtained by studying the solutions for the fields $\chi$, $\Phi$ and $g_{rr}$ as we vary the dimension $d$. In figure \ref{d7171} we show an example of such a solution with $d=7.171$, slightly below the critical dimension $d_c$. Comparing to figure \ref{chiPhid65}, we see that while the radial size of the solution does not change significantly as we change $d$ from $6.5$ to $7.171$, the magnitude of the fields does. In particular, $\chi(0)$ in figure \ref{d7171} is very close to the critical value $1/\sqrt2\pi\simeq 0.22$, and the difference $\Phi(0)-\Phi(\infty)$ is around $2$, which means that the string coupling $g_s\sim e^\Phi$ at small $r$ is larger than that at large $r$ by a factor of $\sim e^{2}$. That is consistent with the fact that $\Phi(0)-\Phi(\infty)\to\infty$ as $d\to d_c$.
\begin{figure}
	\centering
	\subfigure[]{
	\begin{minipage}[t]{0.45\linewidth}
	\centering
	\includegraphics[width=2.5in]{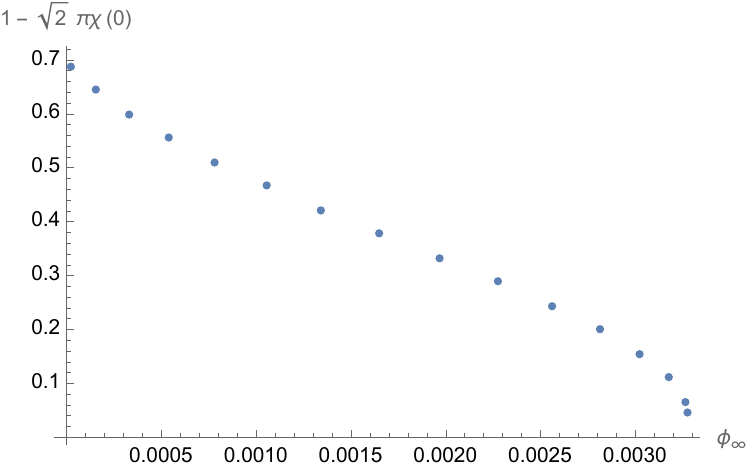}\label{chi0d65}
	\end{minipage}}
	\subfigure[]{
	\begin{minipage}[t]{0.45\linewidth}
	\centering
	\includegraphics[width=2.5in]{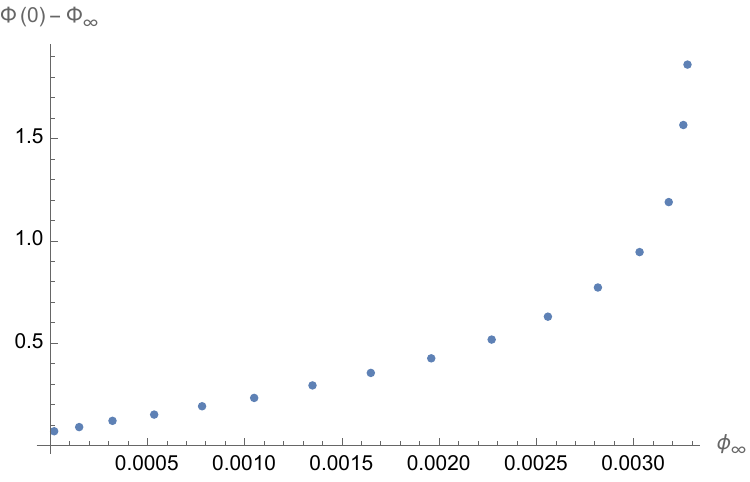}\label{Phi0d65}
	\end{minipage}}
	\centering
\caption{\label{d65}The dependence of $\chi(0)$ and $\Phi(0)$ on $\phi_\infty$ for $d=6.5$.}
\end{figure}

So far, we discussed the solutions of equations \eqref{eh} -- \eqref{cc} that preserve a diagonal $SU(2)$ subgroup of $SU(2)_L\times SU(2)_R$, i.e. the large $k$ analog of the solutions with $\beta=\beta_H$ in \cite{Balthazar:2022hno}. We next describe the solutions that break the symmetry to a diagonal $U(1)$, the large $k$ analogs of solutions with $\beta>\beta_H$. These solutions have the property that the radion $\phi(r)$ does not go to zero at infinity. As discussed in \cite{Chu:2025boe}, its asymptotic value, $\phi_\infty=\lim_{r\to \infty}\phi(r)$ can be thought of as parameterizing $\beta$. 

To see the effect of a non-zero $\phi_\infty$ on the solutions, we fix the dimension, and study the solutions as a function of $\phi_\infty$. In figure \ref{d65} we exhibit the results of this analysis for $d=6.5$. In panel (a) we plot the dependence of $\chi(0)$ on $\phi_\infty$. We see that as $\phi_\infty$ increases (i.e. the temperature decreases), $\chi(0)$ increases, and eventually, for a particular value of $\phi_\infty$, roughly $0.0033$ for $d=6.5$, it approaches the critical value $1/\sqrt2\pi$. In figure \ref{Phi0d65} we plot the behavior of the dilaton, $\Phi(0)-\Phi(\infty)$ as we change $\phi_\infty$. We find that as the latter approaches the critical value, this difference diverges, and beyond that point the solution ceases to exist, as before. 

\begin{figure}
	\centering
	\subfigure[]{
	\begin{minipage}[t]{0.45\linewidth}
	\centering
	\includegraphics[width=2.5in]{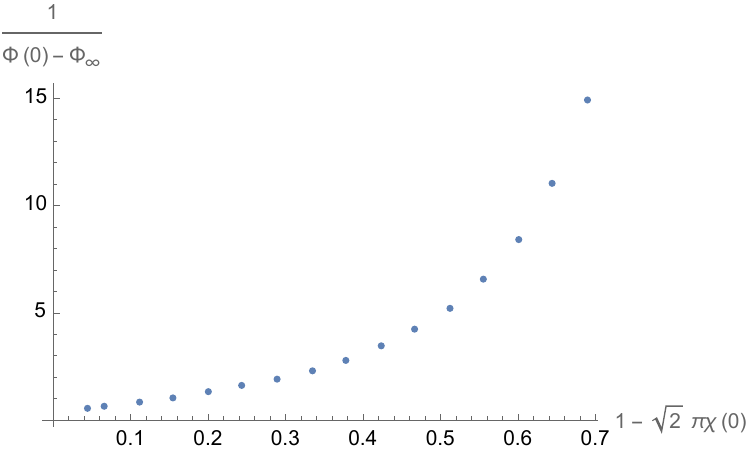}\label{Phi0vschi0d65}
	\end{minipage}}
	\subfigure[]{
	\begin{minipage}[t]{0.45\linewidth}
	\centering
	\includegraphics[width=2.5in]{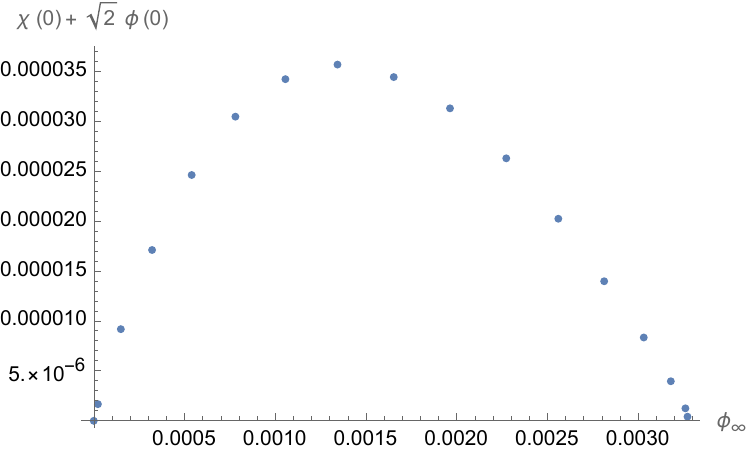}\label{chiphid65}
	\end{minipage}}
	\centering
\caption{\label{mored65}For $d=6.5$, the geometry of the three-sphere and value of the dilaton develop a singularity at $r=0$, at $\phi_\infty\approx 0.0033$.}
\end{figure}
Figure \ref{mored65} provides further information about the approach to the singularity in figure \ref{d65}. In panel (a) we eliminate $\phi_\infty$ and plot the relation between the field $\chi$ and the dilaton at $r=0$. The fact that the resulting curve approaches the origin is an indication of the fact that the dilaton at the origin diverges at the same point at which $\chi(0)$ goes to its critical value.  

In figure \ref{chiphid65} we plot the variable $\chi(0)+\sqrt2\phi(0)$ as a function of $\phi_\infty$. This plot is significant for the following reason. As we discussed before,  $\phi_\infty=0$ corresponds to the Hagedorn temperature, and the HP solution preserves in this case a diagonal $SU(2)$ subgroup of $SU(2)_L\times SU(2)_R$. This means that $\chi+\sqrt2\phi$ vanishes for all $r$. For  $\phi_\infty>0$, the symmetry is further broken to $U(1)$, and one can view the vertical axis in figure \ref{chiphid65} as parametrizing the amount by which the symmetry is broken for small $r$. Figure  \ref{chiphid65} shows that as one approaches the critical value of $\chi$, this amount goes to zero. Thus, the $SU(2)$ symmetry is restored as one approaches the singularity. 

\begin{figure}
	\centering
\includegraphics[scale=0.5]{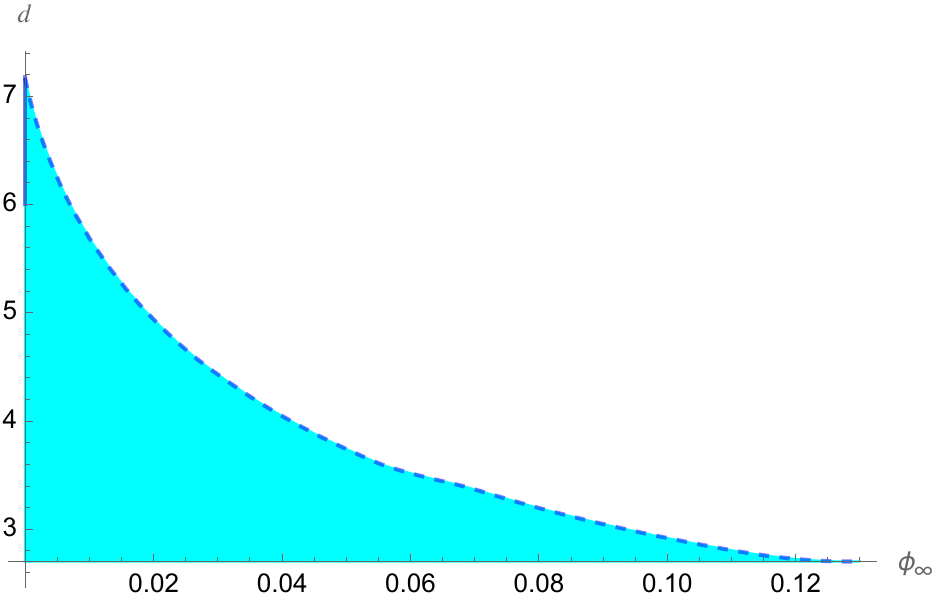}
\caption{\label{dphiinf} HP-like solutions exist in the shaded region. The solid blue line denotes the existence of solutions at $\phi_\infty=0$ for $6<d<d_c\approx 7.171$. The dashed line corresponds to solutions that are singular at $r=0$. }
\end{figure}

Figure \ref{dphiinf} provides a useful summary of the numerical solutions that we found. HP-type solutions exist in the shaded region in that figure. For given $2< d< d_c$ there is a finite range of $\phi_\infty$ in which these solutions exist. The size of this range goes to zero as $d\to d_c$ where, as we saw before, even the solution at $\phi_\infty=0$ ceases to exist. As one approaches the dashed blue line in figure \ref{dphiinf}, the solution becomes singular at small $r$. In particular, the deformed three-sphere becomes singular,  and the dilaton goes to infinity.

\begin{figure}
	\centering
\includegraphics[scale=0.6]{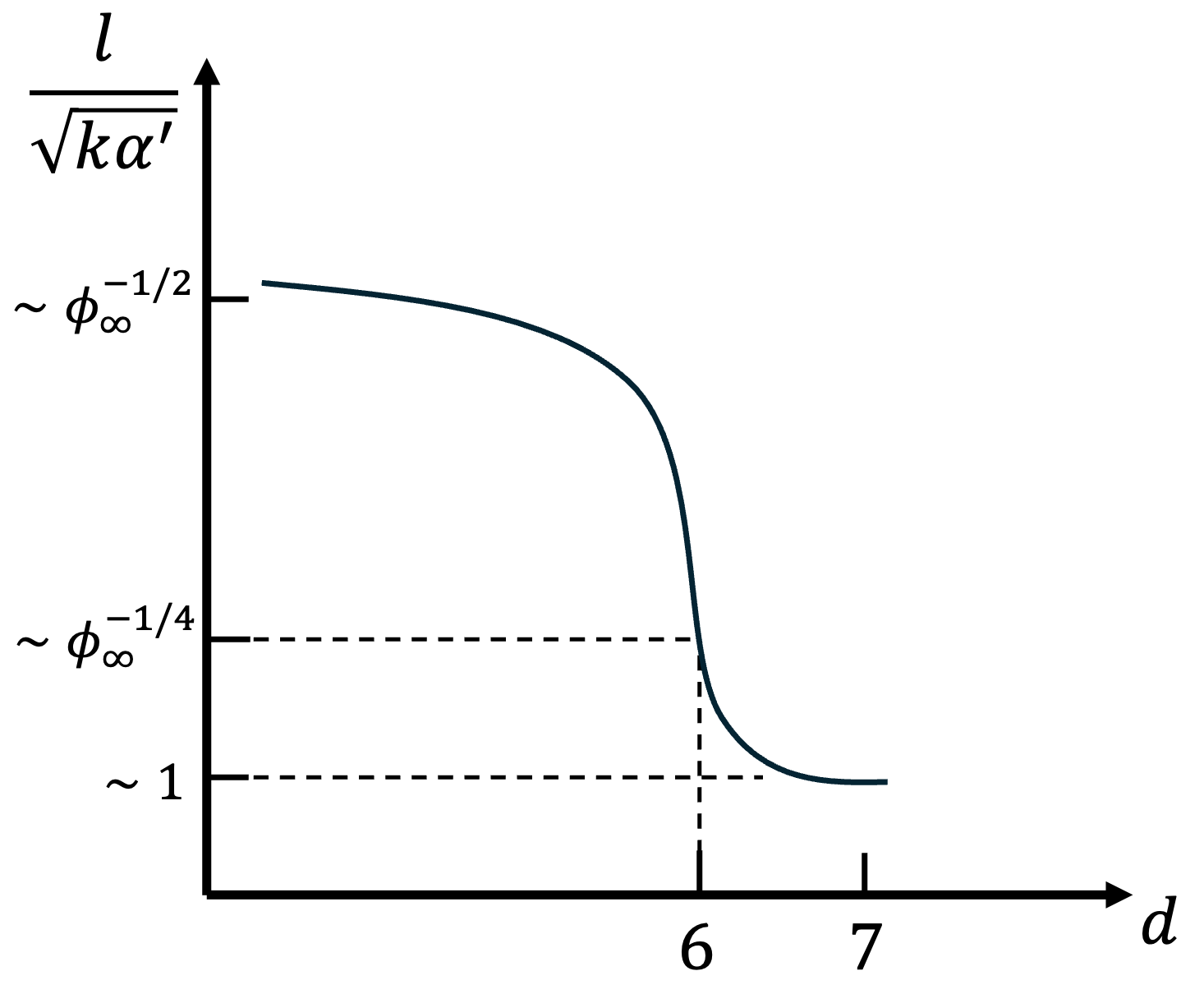}
\caption{\label{ld} Size of HP-like solutions as a function of $d$ at small $\phi_\infty$.}
\end{figure}

It is also instructive to study how the size of the HP-type solutions varies with \(d\). In figure \ref{ld}, we sketch this behavior for a fixed small \(\phi_\infty\). Since the e.o.m. become independent of \(k\) after the coordinate rescaling \eqref{resx}, the natural unit for the size \(l\) is \(\sqrt{k\alpha'}\); in this sense, the large \(k\) limit acts as a magnifying glass. 

Another small parameter in the problem is 
$\phi_\infty\equiv\phi(r\to\infty)$.
Plugging it into the Lagrangian (\ref{kinterm}), \eqref{Vchiphi}, (\ref{resx}), we see that this parameter determines the mass of the winding tachyon $\chi$ at infinity 
\begin{equation}
    m_\infty=\sqrt{\frac{16\pi}{k\alpha'}\phi_\infty}\ .
\end{equation}
For $\phi_\infty\ll1$, the winding tachyon is light, $m_\infty\ll m_s$, and one may expect the size of the solution to go like $1/m_\infty$. As seen in figure \ref{ld}, this is indeed the case for $d<6$, as originally found in \cite{Horowitz:1997jc}. Interestingly, for $d>6$ one finds that the size of the solution does not grow as $\phi_\infty\to 0$. Rather it remains of order $\sqrt{k}l_s$ all the way to $\phi_\infty=0$, the large $k$ analog of $\beta=\beta_H$.  

It is natural to ask what happens when we try to extend our large $k$ analysis to $k$ of order one (where \(\mathbb{S}^3\) is replaced by \(\mathbb{S}^1\)). As we saw before, for $d\le 6$, as well as $d=6+\epsilon$ with $\epsilon\ll 1$, there are two sources of suppression of uncontrolled corrections to the effective action, coming from large $k$ and small $\phi_\infty$, respectively. In this regime, one can extend our analysis to $k$ of order one, the case discussed in \cite{Balthazar:2022szl,Balthazar:2022hno}. On the other hand, in regimes where our analysis is reliable only due to the large $k$ approximation, the original problem becomes uncontrolled using present techniques, as discussed in~\cite{Chu:2025boe}. However, it is natural to expect that the qualitative structure of the solutions is similar to what we found.

\section{Geometry}\label{sec:geometry}

As mentioned in~\cite{Chu:2025boe}, the fields $\phi$ and $\chi$ in (\ref{resclagdil}) can be thought of as geometric deformations of a large three-sphere, supported by $H$-flux~\cite{Witten:1983ar}. In this section we describe these deformations to first order in the fields. 

The sigma model on the three-sphere is described  by the Lagrangian
\begin{equation}
\label{sigma}
    S_{\sigma}=\frac{k}{4\pi}\int d^2z\, {\rm Tr}\, \partial g\bar\partial g^{-1}\ ,
\end{equation}
where $g$ denotes the $SU(2)$ group element, and $k$ is related to the radius of the three-sphere in string units, $R_3\sim\sqrt k$. This can be seen by parametrizing $g$ as
\ie
\label{gX}
g=\begin{pmatrix}
    \hat X_0+i\hat X_3&\ i\hat X_1+\hat X_2\\
   i\hat X_1-\hat X_2&\  \hat X_0-i\hat X_3
\end{pmatrix}\ ,\quad \det g =\hat X_0^2+\hat X_1^2+\hat X_2^2+\hat X_3^2=1\ .
\fe
The action (\ref{sigma}) is manifestly invariant under $SU(2)_L\times SU(2)_R$, which acts on $g$ as $g\to U_L gU_R$, and on the coordinates $\hat X_i$ in \eqref{gX} as rotations on a unit three-sphere.

The unit sphere can be parametrized by the angular coordinates $(\psi,\theta,\varphi)$: 
\begin{equation}
\label{Xtheta}
    \begin{pmatrix}
        \hat X_0\\\hat X_1\\\hat X_2\\\hat X_3\\
    \end{pmatrix}=\begin{pmatrix}
        \cos\psi\\\sin\psi\sin\theta\cos\varphi\\\sin\psi\sin\theta\sin\varphi\\\sin\psi\cos\theta
    \end{pmatrix}\ .
\end{equation}
Substituting (\ref{gX}), (\ref{Xtheta}) into (\ref{sigma}) leads to 
\begin{equation}
    S_{\sigma}=\frac{k}{2\pi}\int d^2z \left(\partial\psi\bar\partial\psi+\sin^2\psi\partial\theta\bar \partial\theta+\sin^2\psi\sin^2\theta\partial\varphi\bar \partial\varphi\right)~,
\end{equation}
from which one can read off the metric on the target space,
\begin{equation}
\label{metS3}
    ds^2=k\alpha'(d\psi^2+\sin^2\psi d\theta^2+\sin^2\psi\sin^2\theta d\varphi^2)\ ,
\end{equation}
the round metric on an $\mathbb{S}^3$ of radius $\sqrt{k}l_s$.

The $SU(2)$ WZW Lagrangian contains another term, corresponding to the $H$-flux. In terms of the fields $(\psi,\theta,\varphi)$, this term can be written as
\begin{equation}
    k\Gamma=\frac{k}{\pi}\int d^2z\,\varphi\sin^2\psi\sin\theta\left(\partial \psi\bar\partial\theta-\partial \theta\bar\partial\psi\right)\ .
\end{equation}
The corresponding $B$-field is given by
\begin{equation}
\label{Bfield}
    B_{\psi\theta}=2k\alpha' \varphi\sin^2\psi\sin\theta \ .
\end{equation}
As discussed in \cite{Chu:2025boe}, the fields $\chi$, $\phi$ correspond in the worldsheet theory to the non-abelian Thirring perturbations
\begin{equation}
\label{perL}
\mathcal{L}_{\rm int}= \lambda_1\mathcal{J}^1\bar{ \mathcal{J}}^1+\lambda_2\mathcal{J}^2\bar{ \mathcal{J}}^2+\lambda_3\mathcal{J}^3\bar{ \mathcal{J}}^3\ ,
\end{equation}
where the $SU(2)$ currents are defined by
\begin{equation}
\label{calJ}
    \mathcal{J}=\begin{pmatrix}
        \mathcal{J}^3&\mathcal{J}^1-i\mathcal{J}^2\\
        \mathcal{J}^1+i\mathcal{J}^2&-\mathcal{J}^3
    \end{pmatrix}\equiv g^{-1}\partial g\ ,\quad \bar{\mathcal{J}}=\begin{pmatrix}
        \bar{\mathcal{J}}^3&\bar{\mathcal{J}}^1-i\bar{\mathcal{J}}^2\\
        \bar{\mathcal{J}}^1+i\bar{\mathcal{J}}^2&-\bar{\mathcal{J}}^3
    \end{pmatrix}\equiv -\bar\partial gg^{-1}\ .
\end{equation}
Note that these currents are normalized to $1/2k$ (see e.g. section 15.4 of~\cite{Polchinski:1998rr}). To leading order, the couplings $\lambda_i$ are related to $\chi$ and $\phi$ by the relations $\lambda_1=\lambda_2=\sqrt2 k\chi$ and $\lambda_3=-2k\phi$~\cite{Chu:2025boe}.

Plugging (\ref{gX}) and (\ref{Xtheta}) into (\ref{calJ}) gives an expression of the perturbation (\ref{perL}) at leading order in $\{\lambda_i\}$. In general, the perturbation (\ref{perL}) breaks the $SU(2)_L\times SU(2)_R$ symmetry. For the case where the diagonal $SU(2)$ is preserved, i.e. $\lambda_1=\lambda_2=\lambda_3=\lambda$, we have
\begin{equation}
\label{aaaa}
\begin{split}
\mathcal{L}_{\rm int}= \frac{\lambda}{2}\Tr \mathcal{J}\bar{\mathcal{J}}=
\lambda\left[\partial\psi\bar\partial\psi+\cos 2\psi\sin^2\psi\left(\partial\theta\bar \partial\theta+\sin^2\theta\partial\varphi\bar\partial\varphi\right)\right.\\
\left.+2\sin^3\psi\cos\psi\sin\theta(\partial\theta\bar\partial\varphi-\partial\theta\bar\partial\varphi)\right]\ .    
\end{split}
\end{equation}
This means that the leading-order perturbations to the background metric (\ref{metS3}) and the $B$-field (\ref{Bfield}) are
\begin{equation}
\label{met1}
    \delta ds^2=2\pi\alpha'\lambda\left[ d\psi^2+\cos2\psi\sin^2\psi\left( d\theta^2+\sin^2\theta d\varphi^2\right)\right]\ ,
\end{equation}
\begin{equation}
    \delta B_{\theta\varphi}=4\pi\alpha'\lambda\sin^3\psi \cos\psi \sin\theta \ ,
\end{equation}
respectively. The residual diagonal $SU(2)$ symmetry preserved by \eqref{aaaa} acts in the usual way on the two-sphere labeled by the Euler angles $(\theta,\varphi)$. 

Using the metric perturbation (\ref{met1}), we can calculate the volume of the deformed three-sphere:
\begin{equation}
\begin{split}    \mathcal{V}_3=&\int_0^{\pi}d\psi\int_0^{\pi}d\theta\int_0^{2\pi}d\varphi\sqrt g\\
=&l_s^3\int_0^{\pi}d\psi\int_0^{\pi}d\theta\int_0^{2\pi}d\varphi\  k^{\frac{3}{2}}\left(1+\frac{\pi\lambda}{k}(1+2\cos 2\psi)+O(\lambda^2)\right)\sin^2\psi\sin\theta\\
=&l_s^3k^{\frac{3}{2}}2\pi^2+O(\lambda^2)\ .
\end{split}
\end{equation}
Thus, to leading order in $\lambda$, the volume does not change. Furthermore, the integral of $\delta H_{\psi\theta\varphi}=\partial_\psi\delta B_{\theta\varphi}$,
\begin{equation}
\delta H_{\psi\theta\varphi}=4\pi \alpha' \lambda \sin\theta (3\sin^2\psi\cos^2\psi-\sin^4\psi)
\end{equation}
over $(\psi,\theta,\varphi)$ also vanishes. This is consistent with the fact that the $H$-flux is quantized.

The $SU(2)$ preserving solutions we found have $\chi=-\sqrt2\phi>0$. Thus,  $\lambda=\sqrt2 k\chi$ is positive. Looking back at (\ref{met1}), we see two kinds of deformations of the metric, to first order in the coupling. First, a positive $\lambda$ implies $\delta g_{\psi\psi}>0$, which means that the size of the $\psi$ dimension increases under this perturbation. Second, the squared radius of the two-sphere labeled by $(\theta,\varphi)$ increases by $2\pi\alpha'\lambda\cos 2\psi \sin^2\psi$. This increase is smaller than the one that would maintain a round three-sphere.

To determine the relation between the couplings $\chi$ and $\phi$ and the geometry of the three-sphere beyond first order in the couplings, it is useful to compare with~\cite{Cvetic:2000dm}, which gives the metric under finite massless deformations. Interestingly, one finds that the critical value $\chi=-\sqrt2 \phi=1/\sqrt{2}\pi$, where the Lagrangian \eqref{resclagdil}-\eqref{Vchiphi} becomes singular, corresponds to a singular metric of the deformed three-sphere~\cite{HPBKL}. In particular, $g_{\psi\psi}$ diverges, while $g_{\theta\theta}$ and $g_{\varphi\varphi}$ vanish near this singularity.

The more general case  $\chi\neq -\sqrt{2}\phi$ corresponds to $\lambda_1=\lambda_2\neq \lambda_3$ in (\ref{perL}). The perturbed metric reads in this case 
\begin{equation}
\label{met2}
\begin{split}
    \delta ds^2=&2\pi\alpha'\lambda_3\left[(\cos\theta d\psi-\sin\psi\cos\psi\sin\theta d\theta)^2-\sin^4\psi\sin^4\theta d\varphi^2\right]\\
    &+2\pi\alpha'\lambda_1\Big[\sin^2\theta d\psi^2+2\sin\psi\cos\psi\sin\theta\cos\theta d\theta d\psi\\
    +&\sin^2\psi(\cos 2\psi-\cos^2\psi\sin^2\theta)d\theta^2+\sin^2\psi\sin^2\theta(\sin^2\psi\sin^2\theta+\cos 2\psi)d\varphi^2\Big]\ .   
\end{split}
\end{equation}
The two-sphere labeled by $(\theta,\varphi)$ is no longer round, in agreement with the fact that the diagonal $SU(2)$ symmetry is broken to $U(1)$ in this case. This $U(1)$ corresponds to translation invariance along the $\varphi$ circle.

\section{Large $d$}

One of the main open problems in the study of Horowitz-Polchinski solutions and their generalizations, reviewed in \cite{Chu:2025boe}, is the relation of these solutions to Euclidean black holes, particularly in the limit where the Hawking temperature of the black hole approaches the Hagedorn temperature. In order to make progress on this problem, one needs to understand what happens to the EBH solution as the Hawking temperature is raised up to the Hagedorn temperature, and what happens to the HP solution as the temperature deviates from the Hagedorn temperature. As discussed in \cite{Chu:2025boe}, both are difficult problems.  

Interestingly, for large $d$ it was shown in \cite{Emparan:2013xia,Chen:2021emg} that the EBH problem remains under control all the way to the Hagedorn temperature. The basic idea of these papers is that at large $d$ the EBH geometry develops a long throat near the Euclidean horizon, where it looks like the two dimensional $SL(2,\mathbb{R})/U(1)$ black hole, with the level of $SL(2,\mathbb{R})$, $k$, determined by the Hawking temperature. In particular, for $\beta=\beta_H$ the level $k$ is equal to $4\;(2)$ for the bosonic (supersymmetric) case. That level is precisely the one for which the $SL(2,\mathbb{R})/U(1)$ theory has an enhanced $SU(2)$ symmetry. Since the two dimensional black hole is well understood for all $k$, so is the large $d$ EBH.

Given this state of affairs, it is natural to ask what happens as we decrease the dimension $d$ and vary the inverse Hawking temperature $\beta$ of the EBH. It has been argued in \cite{Chen:2021dsw} that the HP solutions are separated from the small EBH ones by a phase transition. Roughly speaking, the two have different topologies -- the former have the topology of a cylinder, while the latter have the topology of a disk. Substantiating this picture in the original HP setting is hard,  partly because of the fact that the HP solutions are not understood beyond the weak coupling regime \cite{Chu:2025boe}. It is natural to ask whether our results shed any light on this problem at large $k$. In this section we will comment briefly on this question. 

In order to do this, we need to generalize the construction of \cite{Emparan:2013xia,Chen:2021emg} from ${\mathbb{R}}^d\times {\mathbb{S}}^1$ to ${\mathbb{R}}^d\times {\mathbb{S}}^3$. To do this, we follow the general idea of \cite{Emparan:2013xia} (the details are somewhat different). We reparametrize the string frame metric on ${\mathbb{R}}^d$, (\ref{dsr}), using the ansatz
\begin{equation}
\label{dsra}
    ds^2=d\rho^2+\rho_0^2e^{\frac{2}{d-1}\Psi(\rho)}d\Omega_{d-1}^2\ ,
\end{equation}
where at this point $\rho_0$ is an arbitrary length scale. Note also that in this parametrization, the radial coordinate $\rho$ varies between $-\infty$ and $+\infty$, in contrast with \eqref{dsr}, where the radial coordinate $r$ varies from $0$ to $\infty$.

The  scalar curvature of \eqref{dsra} is
\begin{equation}
    \mathcal{R}=\frac{(d-1)(d-2)}{\rho_0^2}e^{-\frac{2}{d-1}\Psi(\rho)}-\frac{d}{d-1}\left(\Psi'(\rho)\right)^2-2\Psi''(\rho)\ .
\end{equation}
The dilaton gravity part of (\ref{resclagdil}) takes in this parametrization the form 
\begin{equation}
\label{ldill}
\begin{split}
    L_\text{dg}=&\sqrt{g}e^{-2\Phi}\left[-\mathcal{R}-4(\nabla\Phi)^2\right]\\
    =&\rho_0^{d-1}e^{\Psi}e^{-2\Phi}\left[-\frac{(d-1)(d-2)}{\rho_0^2}e^{-\frac{2}{d-1}\Psi}-\frac{d-2}{d-1}(\Psi')^2+4\Phi'\Psi'-4(\Phi')^2\right]\ ,
\end{split}
\end{equation}
where we have integrated the term involving $\Psi''$ by parts (and neglected the boundary term). Defining the lower dimensional dilaton (after reduction on the $(d-1)$-sphere), $\Phi_1\equiv \Phi-\frac{\Psi}{2}$, the Lagrangian \eqref{ldill} takes the form 
\begin{equation}
   L_\text{dg}= \rho_0^{d-1}e^{-2\Phi_1}\left[-\frac{(d-1)(d-2)}{\rho_0^2}e^{-\frac{2}{d-1}\Psi}+\frac{(\Psi')^2}{d-1}-4(\Phi'_1)^2\right]\ .
\end{equation}
In the region where $\Psi(\rho)\ll d$ (for large $d$), this Lagrangian reduces to
\begin{equation}
\label{onedL}
   L_\text{dg}= -4\rho_0^{d-1}e^{-2\Phi_1}\left[\lambda^2+(\Phi'_1)^2\right]\ ,
\end{equation}
where
\begin{equation}
\label{formlambda}
   \lambda^2=\frac{(d-1)(d-2)}{4\rho_0^2}~
\end{equation}
is an arbitrary constant. 
As in \cite{Emparan:2013xia}, we take $d, \rho_0\to\infty$, while keeping $\lambda=\frac{d}{2\rho_0}$ fixed. The solution of the equation of motion of \eqref{onedL} is a linear dilaton background ${\mathbb{R}}_\rho$, 
\begin{equation}
   \Phi_1=\pm\lambda \rho~.
\end{equation}
This solution is of course singular -- the string coupling diverges at one of the two limits $\rho\to\pm\infty$. 

So far we focused on the ${\mathbb{R}}^d$ part of ${\mathbb{R}}^d\times {\mathbb{S}}^3$. To resolve the singularity mentioned in the previous paragraph, we need to add it back. Thus, the full singular background is ${\mathbb{R}}_\rho\times {\mathbb{S}}^3$. 
Recall that up to this point, the constant $\lambda$ \eqref{formlambda} was arbitrary, but there is a particular value of this constant for which this problem reduces to one that has been solved in a different context, in the supersymmetric version of this theory. This was discussed in section 7 of \cite{Chu:2025boe}, so here we will be brief.  

For $\lambda^2=2/k$, the background ${\mathbb{R}}_\rho\times {\mathbb{S}}^3$ is precisely the near-horizon geometry of $k$ NS5-branes \cite{Callan:1991at}. In that case, the strong coupling singularity can be resolved by separating the NS5-branes, i.e. going to the Coulomb branch of the theory. This (spontaneously) breaks the $SO(4)=SU(2)_L\times SU(2)_R$ symmetry of rotations about the fivebranes to a subgroup, and cuts off the radial direction ${\mathbb{R}}_\rho$ at a finite value of $\rho$. 

Therefore, we conclude that at large $d$, the ${\mathbb{R}}^d\times {\mathbb{S}}^3$ theory has solutions that are analogous to EBH's, in which the radial direction is cut-off in a smooth way at a finite value of the radial coordinate. In terms of figure \ref{dphiinf}, these solutions live at large $d$ and any $\phi_\infty$. It is an interesting open question whether they fill the whole unshaded region in that figure, and are thus separated by a phase transition from the HP solutions, that fill the shaded region.

\section{Discussion}

The main goal of this paper was to complete the analysis of \cite{Chu:2025boe} and solve the equations that describe the large $k$ analogs of (generalized) Horowitz-Polchinski solutions. The result of this analysis is summarized in figure \ref{dphiinf}. Interestingly, these solutions only exist in a small region of the $(d,\beta)$ plane, where $d$ is the dimension of space and $\beta$ the inverse temperature. At the boundary of this region, described by the dashed line in the figure, the kinetic and potential terms for $(\phi,\chi)$ become singular and the dilaton diverges as $r\to 0$.

There are a number of questions that our results raise. 
\begin{itemize}
    \item Is there a qualitative way to understand the form of the shaded region in figure \ref{dphiinf}, without solving the equations? In particular, why are there maximal values of the dimension $d$ and inverse temperature $\beta$ for which solutions exist?
    \item In our analysis, the maximal dimension $d_c$ was obtained numerically. Is there an analytic argument for it? 
    \item Our analysis took place for large $k$. Is the picture similar for the actual HP problem, which has $k=1$ for the bosonic case, and $k=2$ for the supersymmetric one? 
    \item Our main focus was on HP-like solutions. Where are the Euclidean black holes in the description of the analog of figure \ref{dphiinf}?
\end{itemize}
We leave these and other questions to future work.

\section*{Acknowledgements}
JC thanks Ningbo University, Zhejiang University, Shanghai Institute for Mathematics and Interdisciplinary Sciences, and Kavli Institute for the Physics and Mathematics of the Universe for hospitality during the conclusion of this work. DK thanks Tel Aviv University and the Weizmann Institute for hospitality. This work was supported in part by DOE grant DE-SC0009924. The work of JC was further supported in part by DOE grant 5-29073.

\vskip 2cm

\bibliographystyle{JHEP}
\bibliography{HP2}

\end{document}